\documentclass[aps,pre,floatfix,showkeys,amsmath,amssymb,superscriptaddress,10pt,longbibliography,pdftex]{revtex4-1}

\usepackage{amsthm}
\usepackage{bm}
\usepackage{epsfig}
\usepackage{verbatim}
\usepackage{graphicx}
\usepackage{physics}
\usepackage{nicefrac}
\usepackage{caption}
\usepackage{subcaption}
\usepackage{hyperref}

\newcommand{\beq}{\begin{equation}}
\newcommand{\eeq}{\end{equation}}
\newcommand{\beqa}{\begin{eqnarray}}
\newcommand{\eeqa}{\end{eqnarray}}
\newcommand{\bc}{\begin{center}}
\newcommand{\ec}{\end{center}}
\newcommand{\bfig}{\begin{figure}}
\newcommand{\efig}{\end{figure}}

\usepackage{xcolor}

\begin{document}

\title{Finite-size relaxational dynamics of a spike random matrix spherical model}

\author{Pedro H. de Freitas Pimenta}
\affiliation{Universidade Federal Fluminense, Departamento de F\'isica,
Av. Gal. Milton Tavares de Souza s/n, 
  Campus da Praia Vermelha, 24210-346 Niter\'oi, RJ, Brazil}
\author{Daniel A. Stariolo}
\affiliation{Universidade Federal Fluminense, Departamento de F\'isica and 
  National  Institute of Science and Technology for Complex Systems, Av. Gal. Milton Tavares de Souza s/n, 
  Campus da Praia Vermelha, 24210-346 Niter\'oi, RJ, Brazil}

\date{\today}
						
\begin{abstract}
We present a thorough numerical analysis of the relaxational dynamics of the
Sherrington-Kirkpatrick spherical model with an additive non-disordered perturbation for
large but finite sizes $N$.
In the thermodynamic limit and at low temperatures, the perturbation is responsible for a
phase transition from a spin glass to a ferromagnetic phase.
We show that finite size effects
induce the appearance of a distinctive slow regime in the relaxation dynamics,
the extension of which depends on the size of the
system and also on the strength of the non-disordered perturbation. 
The long time dynamics is characterized by the two largest
eigenvalues of a spike random matrix which defines the model, and particularly by the statistics of
the gap between them. We characterize the finite size statistics of the two largest eignevalues of
the spike random matrices in the different regimes, sub-critical, critical and super-critical,
confirming some known results and anticipating
others, even in the less studied critical regime.
We also  numerically characterize the finite size
statistics of the gap, which we hope may encourage analytical work which is lacking. Finally, we compute
the finite size scaling of the long time relaxation of the energy, showing the existence of power laws
with exponents that depend on the strenght of the non-disordered perturbation, in a way which is
governed by the finite size statistics of the gap.
\end{abstract}


\maketitle

\section{Introduction}
Quenched random interactions are known to be at the origin of complex behavior in many body systems,
both in the thermodynamics and dynamics as well~\cite{MePaVi1987,GiardinaDominicis2006,Parisi2023}.
The understanding of the properties of this kind of systems in the
thermodynamic limit has been steadily growing in the last 40 years or so, mainly through 
solutions of mean field, fully connected, models and also from numerical simulations of
finite dimensional ones. On the other side, the behavior of systems composed of a large but finite
number of degrees of freedom is much less understood. This case is relevant in many applications in
many branches of science, e.g. optimization and inference algorithms~\cite{MeMo2009}, biological
populations~\cite{May1973}, neural networks~\cite{Hopfield1982}, to cite but a few.
Powerful techniques, like the saddle point method, are not so
useful for studying systems far from the thermodynamic limit. 
There are a few class of models in which
both the thermodynamics and the dynamical behavior as well can be solved exactly and still they show interesting
non-trivial properties qualitatively similar to more complex systems. Well known examples are systems
in which the degrees of freedom obey a spherical constraint. One of these models with quenched random
pairwise interactions, the Spherical Sherrington-Kirkpatrick model (SSK) allows an exact solution of its
thermodynamic properties using tools from Random Matrix Theory~\cite{KTJ1976}, without the need to use
the more involved replica formalism, necessary for the Ising case. The Langevin relaxational dynamics of the
SSK model was solved in~\cite{Cugliandolo1995}, where it was shown that the long time relaxation is
slow, e.g. with the energy density decaying with a power law in time after a quench from a high temperature
initial state to a temperature below the spin glass transition temperature. Interesting out of equilibrium
features of the dynamics, like the phenomenon of aging, is present in the model and were completely
characterized. The solution of both the thermodynamics and the dynamics of the model were possible due
to the knowledge of the spectral properties of the random interactions matrix. In the case of the
Gaussian Orthogonal Ensemble, the relevant information is in the Wigner semi-circle density of eigenvalues.
When the size of the matrix, $N\times N$, is large but finite the situation changes. The support of
the eigenvalue density is not limited anymore, and the probability distribution of the eigenvalues on
the soft edge is given by the celebrated Tracy-Widom $\beta$ distributions~\cite{TracyWidom1994,TracyWidom1996},
where $\beta=1,2,4$ refer to the orthogonal, unitary and symplectic ensembles, respectively.
Fluctuations of the free energy of the SSK model were studied, e.g.
in~\cite{Baik2016,Baik2017,Johnstone2021,Landon2022}. These fluctuations are
governed by the statistics of the largest eigenvalue of the GOE interaction matrix. The finite $N$
fluctuations in the Langevin dynamics were studied in~\cite{Fyodorov2015,Barbier2021}. In these works, a new
algebraic (power law) scaling regime was found and characterized, not present in the $N\to \infty$ regime.
In the dynamical context, besides the relevance of the largest eigenvalue, which is directly proportional
to the ground state energy, also the gap between the two largest eigenvalues is a fundamental quantity to
compute the relevant time/size scalings of the long time relaxation. 

A related interesting model is the SSK model supplemented with an additive Curie-Weiss term in the Hamiltonian,
or equivalently, where the original random interaction matrix is perturbed by a rank one matrix which has the effect of
shifting the average value of the random matrix elements from zero to a non zero value. The thermodynamics of
this model was also solved in the original work by Kosterlitz et.al.~\cite{KTJ1976}. At low enough
temperatures, the model presents a phase transition from a spin glass to a ferromagnetic phase, at a critical
value of the relative strength between the random interactions and the Curie-Weiss one. In the mathematics
literature, this kind of random matrices with finite rank perturbations are called ``spike random matrices''.
There is a large body of work devoted to the study of the spectral properties of spike random matrices
~\cite{Johnstone2001,Baik2005,Peche2006,Feral2007,Capitaine2012,Mo2012,Bloemental2013,Pizzo2013}.
Of special interest for the physics community is the result, originally presented in~\cite{Baik2005},
of a sharp phase transition in the statistics of the largest eigenvalue of particular classes of spike
random matrices. Of course, this phase transition has an immediate interpretation in the context of the
thermodynamic and also dynamic transitions in the SSK model and related ones.
Recently, a renewed interest in the statistical behavior of spike random matrices is manifested in
several works focusing in different applications, e.g overlaps between eigenvectors of correlated
spike random matrices~\cite{Pacco2023}, analysis of optimal learning rates in non-convex optimization
~\cite{dAscoli2022}, low-rank matrix estimation~\cite{Guionnet2022}, limits of detection of planted
states~\cite{Alaoui2020}, ruggedness of complex energy landscapes~\cite{Ros2019}. From a dynamical
perspective, understanding of spike random matrix models may shed light on problems like the feasibility
of identifying a deterministic signal in a random environment, the reconstruction of hidden patterns
in a complex landscape or the efficency of search algorithms.

In this work we perform a numerical study of the statistics of the largest eigenvalues and
the gap between the two largest ones in spike random matrices from the GOE ensemble. With the information
gained, we then describe the relaxation of the excess energy from the ground state of the spike SSK model
following a quench from a high temperature initial state directly to zero temperature. We show that, as
is the case in the standard SSK model, the relaxation shows a new scaling regime, present when the
system transitions from the spin glass to the ferromagnetic phases, i.e. a critical scaling regime. In
this critical sector, we show that the scaling relaxation behavior is governed by a one parameter scaling
function which depends on the relative strength of the random and Curie-Weiss term and also on the size
of the system.

The paper is organized as follows: in Section \ref{sec:themodel} we introudce the model
studied and summarize some known results which will be useful later; in Section \ref{sec:eigenstatistics}
we present a numerical study of the statistics of the two largest eigenvalues and the gap of a rank one spike GOE
matrix; in Section \ref{sec:dynamics} we present our results on the relaxation dynamics of the model,
both in the thermodynamic limit and for large but finite system sizes. Finally, in Section \ref{sec:conclusions},
we make a brief discussion of the work and present our conclusions.
\section{The model} \label{sec:themodel}

The spherical Sherrington-Kirkpatrick (SSK) model with a Curie-Weiss (CW) perturbation is described by
the following Hamiltonian:
\begin{equation}
\begin{aligned}
    \mathcal{H}[\vec{S},z] &=
    \mathcal{H}_{SSK}(\vec{S}) +
    \mathcal{H}_{CW}(\vec{S})+\frac{z}{2}\left(\vec{S}\,^2-N\right)\\
    &= -\frac{1}{2}\vec{S}\,^T\mathbf{M}\vec{S}+\frac{z}{2}\left(\vec{S}\,^2-N\right)\\
    &=-\frac{1}{2\sqrt{N}}\sum_{i\neq j}^N J_{ij}\, s_i s_j-\frac{\theta}{2N}\sum_{i\neq j}^N\, s_i s_j
    + \frac{z}{2}\left(\sum_i^N s_i^2 -N\right),\\
\label{eq:CW_perturb}
\end{aligned}
\end{equation}
where $z$ is a Lagrange multiplier which enforces the spherical constraint:
\beq
\vec{S}\,^2 = \sum_i^N s_i^2 = N.
\label{eq:spherical-constraint}
\eeq
In the previous expressions, the spin variables $ s_i \in [-\sqrt{N}, \sqrt{N}]$ are described by a $N$ component vector
$\Vec{S}=(s_1,...,s_N)$. The coupling constants $J_{ij}$ are chosen form a real and
symmetric random matrix from the Gaussian Orthogonal Ensemble (GOE), $\boldsymbol{J}=\{J_{ij}\}_{(i,j)\in[1,N]^2}$,
with zero mean and variance $J^2$. $\theta\in\mathbb{R^+}$ measures the intensity of the deterministic perturbation.
Thus, $\boldsymbol{M}$ is a real symmetric $N\times N$ spike random matrix, whose off-diagonal elements are Gaussian distributed
with the following mean and variance:
\begin{equation}
    p(M_{ij}) = \mathcal{N}\left[\mu=\frac{\theta}{N},\;\sigma^2=\frac{J^2}{N}\right]\;.
    \label{spikematrices-entries}
\end{equation}
The diagonal elements are zero.
The Hamiltonian can be rewritten decomposing $\Vec{S}$ as a linear combination of the eigenvectors $\{\Vec{V}_\mu\}$ of
the coupling matrix $\boldsymbol{M}$, with $\Vec{V}_\mu\cdot\Vec{V_\nu}=\delta_{\mu\nu}$. Therefore, with the following
notation $s_\mu=\Vec{S}\cdot\Vec{V}_\mu $ for the projections of $\Vec{S}$ on the eigenvectors of $\boldsymbol{M}$,
the Hamiltonian becomes:
\begin{equation}
    \mathcal{H}[\Vec{S},z]=
    -\frac{1}{2}\sum_{\mu=1}^N (\lambda_\mu-z)\,s_\mu^2-\frac{z}{2}N\;, 
    \label{HSSK-energy}
\end{equation}
with $\{\lambda_\mu\}_{\mu\in[1,N]}$ being the set of $N$ eigenvalues of $\boldsymbol{M}$, with associated eigenvectors
$\{\vec{V}_\mu\}$. The eigenvalues are organised such that
$\text{Max}[{\lambda_\mu}]=\lambda_1>\lambda_2>...>\lambda_N=\text{Min}[{\lambda_\mu}]$.
In the large $N$ limit, the eigenvalue density distribution of $\boldsymbol{M}$ is given by~\cite{EdwardsJones1976}: 
\begin{equation}
    \rho(\lambda)=
    \begin{cases}
         \rho_W(\lambda)\;\;&,\;\; \theta \leq J \\
         \rho_W(\lambda)+\frac{1}{N}\delta\left[\lambda-\left(\theta+J^2/\theta\right)\right]\;\;&,\;\; \theta>J
    \end{cases}\;,
\label{eq:spike-spectrum}
\end{equation}
where $\rho_W(\lambda)$ is the Wigner semicircle law:
\begin{equation}
    \rho_W(\lambda)=
    \begin{cases}
         \frac{\left(4J^2-\lambda^2\right)^{1/2}}{2\pi J^2}\;\;
         &,\;\;|\lambda|<2J \\
         0\;\;&,\;\;|\lambda|>2J
    \end{cases}\;.
    \label{WignersLaw}
\end{equation}
The result (\ref{eq:spike-spectrum}) means that, if $\theta \leq J$, the spectrum of $\boldsymbol{M}$ is
given by the Wigner law, corresponding to the GOE ensemble.
Otherwise, if $\theta > J$, the largest eigenvalue $\lambda_1$ detaches from the Wigner semicircle,
becoming an outlier with a delta peak at $\lambda=\theta+J^2\theta^{-1}$.

The overdamped dynamics of the model is governed by the set of Langevin equations :
\begin{equation}
\begin{aligned}
    \frac{\partial s_i(t)}{\partial t}
    & = -\delta_{s_i} \mathcal{H}[\Vec{S},z] + \xi_i(t) \\
    & = \sum_{j\neq i}^N M_{ij}s_j(t)-z(t,\{s_\mu(0)\})s_i(t)+\xi_i(t),
    \;\;\;\;\forall i\in[1,N]\;,   
\end{aligned}
\end{equation}
where $\xi_i(t)$ represents a Gaussian white noise with zero mean and variance
$\langle \xi_i(t)\xi_i(t')\rangle=2T\delta_{ij}\delta(t-t')$ and $T$ is the temperature of a thermal bath.
As in \cite{Barbier2021}, here we are interested in the zero temperature limit of the Langevin equations which,
in the eigenbasis of $\boldsymbol{M}$, reads:
\begin{equation}
    \frac{\partial s_\mu(t)}{\partial t}=\left[\lambda_\mu-z(t,\{s_\mu(0)\})\right]s_\mu(t), \;\;\;\;\forall \mu\in[1,N]\;.
    \label{eq:Langevin-eqs}
\end{equation}
At long times, the system must fall in a stable or metastable  state of the free energy which, at $T=0$,
reduces to the Hamiltonian $\mathcal{H}[\vec{S},z]$. The asymptotic stationary state will depend on the initial conditions
$\{s_\mu(0)\}$, as stated explicitely in (\ref{eq:Langevin-eqs}).
By setting $\lim_{t\to\infty} \partial_t s_\mu(t) =0$ we obtain the criteria:
\begin{equation}
    \delta_{s_\mu}\mathcal{H}[\vec{S},z]=-(\lambda_\mu-z)s_\mu=0, \quad\qquad \forall \mu \in [1,N],
\end{equation}
complemented by the spherical constraint in eq.~(\ref{eq:spherical-constraint}). This system of equations admits the $2N$
solutions:
\begin{equation}
\vec{S}=\pm\sqrt{N} \,  \vec{V}_\mu \qquad\qquad \mbox{and}  \qquad\qquad z=\lambda_\mu \quad\qquad \forall \mu \in [1,N].
\end{equation} 
Their stability is determined by the Hessian $\delta_{s_\mu}\delta_{s_\nu} \mathcal{H}[\vec{S},z]=-(\lambda_\nu-z)\delta_{\mu \nu}$. 
Taking a given metastable state $\vec{S}=\sqrt{N}\vec{V}_\mu$, the local landscape has $N-\mu$ stable directions,
$\mu-1$ unstable directions and a marginal flat one. 
The energy of each of these configurations is equal to $- \lambda_\mu N/2$. Thus, the system should always equilibrate
in one of the solutions $\pm\sqrt{N}\vec{V}_1$, as they are the only stable ones. The ground state energy density is then
simply given by $e_{\rm eq} = - \lambda_1/2$.

Our primary interest here is to describe the behaviour, at long times, of the excess energy density,
$\Delta e(t,N)=e(t,N)-e_{\rm eq}(N)$, for arbitrary system sizes $N$. To this end, recalling results in
\cite{Fyodorov2015,Barbier2021}, it can be shown that the time dependent Lagrange multiplier has a simple relation with the energy
density, $z(t,N) = -2e(t,N)$, leading to the exact expression for the excess energy density:
\begin{equation}
  \Delta e(t,N)=\frac{\lambda_1}{2}+e(t,N)=
  \frac{1}{2}\frac{\sum_{\mu=2}^N s_\mu^2(0)(\lambda_1-\lambda_\mu)\,e^{2(\lambda_\mu-\lambda_1)t}}
       {s_1^2(0)+\sum_{\mu=2}^N s_\mu^2(0)\,e^{2(\lambda_\mu-\lambda_1)t}}\;.
   \label{eq:energy excess}
\end{equation}
Because of the dependence of the above expression on the relaxation rates $\lambda_\mu-\lambda_1$, one expects that the late
dynamics of the model will be dominated by the gap,
$g=\lambda_1-\lambda_2$, between the two largest eigenvalues of the random matrix $\boldsymbol{M}$:
\begin{equation}
    \Delta e(t,N)\xrightarrow{t\to\infty}\frac{1}{2}\frac{s_2^2(0)}{s_1^2(0)}\,g\,e^{-2gt}\;.
\end{equation}
In the previous expression there are two sources of fluctuations: the statistics of the gap and the initial conditions.
The vector of initial conditions $\vec{S}(0)=\{s_\mu(0)\}$ can be 
written in the basis of eigenvectors of the $\boldsymbol{M}$ matrix in the form $\vec S(0) = \sum_\nu c_\nu \vec V_\nu$,
where the coefficients $c_\nu$ statisfy the condition
$S^2(0) = \sum_{\nu\eta} c_\nu c_\eta \, \vec V_\nu \cdot \vec V_\eta = \sum_{\nu} c^2_\nu = N$.
In the present work, we are primarily interested in a flat distribution on the basis of eigenvectors,
that is $c_\nu=1$ for all $\nu$, which can be associated to 
thermal equilibrium at a very high temperature. It corresponds to: 
\begin{equation}
  s_\mu(0) = s^{\rm flat}_{\mu}(0)=1 \quad\qquad \forall \mu \in [1,N].
\label{eq:rdminitcond}  
\end{equation}
The relaxation dynamics of the model depends on the form of the eigenvalue density (\ref{eq:spike-spectrum}).
The effect of the delta contribution is to induce a phase transtion when the intensity of the Curie-Weiss term
attains the value $\theta = J$ in the thermodynamic limit. While the CW term remains weaker
than the random couplings intensity, $\theta<J$, the system behaves like the pure SSK model,
relaxing towards a disordered ground state $\vec S$ with a characteristic slow dynamics, as described in
\cite{Cugliandolo1995,Fyodorov2015,Barbier2021}. At finite temperatures, the thermodynamics corresponds to a spin glass phase,
originally described in~\cite{KTJ1976}.
On the other hand, if the perturbation is strong enough, $\theta>J$, the largest eigenvalue detaches from the bulk
of the spectrum, inducing a fast relaxation towards a ferromagnetic ground state, where all the spin variables $s_i$
align in the same direction. For not too high temperatures, exactly at $\theta =J$, the system goes through a continuous
phase transtion between a disordered spin glass phase and a ferromagnetically ordered one, in the thermodynamic limit
~\cite{KTJ1976}.
In~\cite{Baik2017} finite size fluctuations of the free energy of the model at both sides of the spin glass-ferromagnetic
transition where characterized. Here, we are
interested in characterizing the finite size fluctuations of the relaxation dynamics, following a quench from an infinite
temperature initial state down to zero temperature, for different values of the CW perturbation intensity.

Considering random initial conditions as givem by (\ref{eq:rdminitcond}), at long times, the behavior of the average
excess energy is given by:
\begin{equation}
\mathbb{E}[\Delta e(t,N)]\xrightarrow{t\to\infty}\frac{1}{2}\,\mathbb{E}[g\,e^{-2gt}]\;.
\label{eq:avexenergy}
\end{equation}
At present, the statistical properties of the gap $g=\lambda_1-\lambda_2$ are not known. In order to describe its approximate
behaviour, in the following we will pursue a thorough numerical investigation of the statistics of the two largest
eigenvalues and the gap of the spike random matrix $\boldsymbol{M}$, for large but finite system size $N$.
\section{Statistics of the two largest eigenvalues and the gap for finite size spike matrices}
\label{sec:eigenstatistics}
\begin{figure}
    \centering
    \includegraphics[width=.8\textwidth]{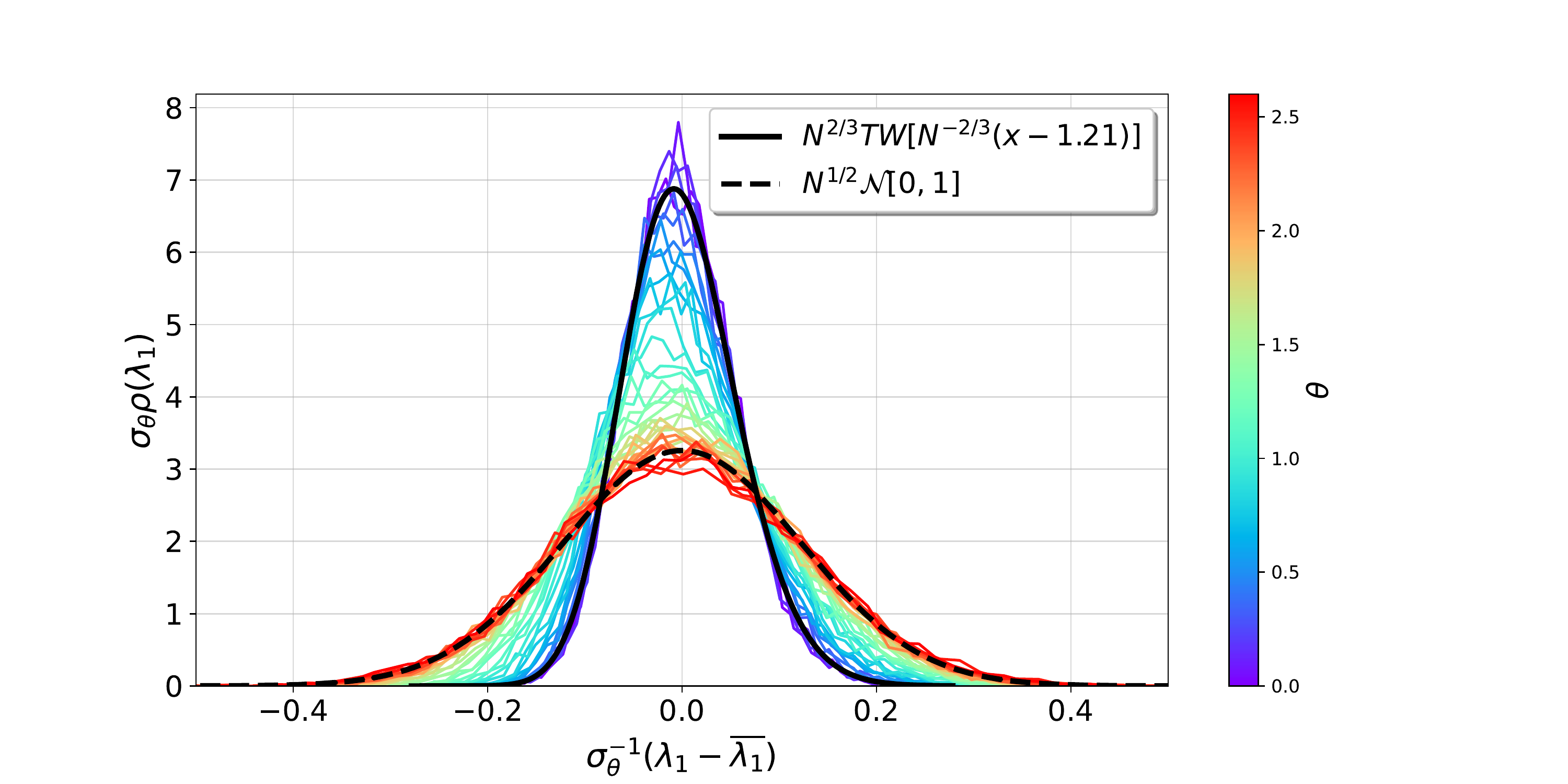}
    \caption{Scaled probability density distributions of an ensemble of $10^4$ spike random matrices with $N=100$.
The distributions are centered relative the ensemble average $\overline{\lambda_1}$ and $\sigma_\theta$ stands
for the predicted standard deviation when $\theta>1$. The centered TW distribution $TW$ (\ref{eq:spike_l1_tw})
and the normal distribution $\mathcal{N}[0,1]$ (\ref{eq:spike_l1_gauss}) have been scaled similarly to the data.}
    \label{fig:dist_l1_spike_N=100}
\end{figure}

In the limit $N \to \infty$ the distribution of eigenvalues is given by (\ref{eq:spike-spectrum}). When $N$
is finite, the border of the spectrum shows finite size fluctuations. For spike matrices belonging to the complex
Wishart ensemble a phase transition was identified in the behavior
of $\lambda_1$, as the mean value of the elements of the random matrix changes~\cite{Baik2005,Peche2006}.
A similar behavior for real Wishart matrices was conjectured in~\cite{Baik2005} and subsequently confirmed by
several approaches (see e.g. \cite{Mo2012,Bloemental2013} and references therein). Extensions for the GOE and other
Gaussian ensembles were considered in~\cite{Bloemental2013}.
Its connection with the thermodynamic phase transition in the SSK model is immediate because the free energy
of the model (which reduces to the average Hamiltonian at zero temperature) is proportional to $\lambda_1$.
Then, when considering finite size fluctutations at $T=0$, three regimes are of interest: a sub-critical regime
when $\theta \ll J$, a critical one when $\theta \sim J$ and a super-critical one when $\theta \ll J$. The fluctuations of
$\lambda_1$ in the sub-critical and super-critical regimes have been considered in several works
~\cite{Feral2007,Capitaine2012,Pizzo2013,Baik2016,Baik2017}. Nevertheless, results on the critical regime are
scarce~\cite{Mo2012,Bloemental2013}.
The following results are known: fixing $J=1$, as long as $\theta<1$, the perturbation has little effect on the behavior of
$\lambda_1$. In this case its distribution is described by the GOE Tracy-Widom (TW) distribution
~\cite{TracyWidom1996,Pizzo2013,Baik2017}:
\begin{equation}
N^{\nicefrac{-2}{3}}\rho\left[N^{\nicefrac{2}{3}}(\lambda_1 - \mathbb{E}[\lambda_1])\right]\Longrightarrow TW,    \hspace{2cm} \theta<1\;.
    \label{eq:spike_l1_tw}
\end{equation}
Instead, when $\theta>1$, $\lambda_1$ becomes an isolated eigenvalue as it goes away from the support of the
semicircle, being freer to fluctuate around the expected value, $\mathbb{E}[\lambda_1]=\theta+\theta^{-1}$.
In this case, the fluctuation of $\lambda_1$ is of order $\mathcal{O}(N^{-1/2})$, described by a normal distribution
~\cite{Pizzo2013,Baik2017}:
\begin{equation}
    N^{\nicefrac{-1}{2}}\rho\left[N^{\nicefrac{1}{2}}(\lambda_1-\mathbb{E}[\lambda_1])\right]
    \Longrightarrow
    \mathcal{N}\left[0,\;2(1-\theta^{-2})\right], \hspace{1cm} \theta>1\;.
    \label{eq:spike_l1_gauss}
\end{equation}
Figure \ref{fig:dist_l1_spike_N=100} shows the behavior of the probability density distribution
of $\lambda_1$,  collected from an ensemble of $10^4$ spike random matrices of size $N=100$, for several values
of $\theta$, shown in color scale to the right of the figure. The distributions
are centered at zero, $\overline{\lambda_1}$ stands for the ensemble average and $\sigma_\theta$ is the predicted
standard deviation given in (\ref{eq:spike_l1_gauss}), applied only in its valid interval $\theta>1$.
The Tracy-Widom and the normal distribution, shown in continuous and dashed lines, are properly scaled. The plots
of the TW distributions were done by using the  publicly available package in 
https://github.com/yymao/TracyWidom/. This package uses interpolation tables from~\cite{BOROT2012,Bejan2006}. 
In agreement with the results above, for $\theta\ll 1$ the pdf of the largest eigenvalue is well described by
a TW distribution. At the other end, when $\theta \gg 1$ a normal distribution with the theoretically predicted
behavior is observed. It is also observed a crossover behavior at intermediate values of $\theta$. This is the
critical regime. At present, there are a few results on the behaviour of the largest eigenvalue of spike random
matrices in the critical regime~\cite{Mo2012,Bloemental2013}, from which we have been able to describe the scaling of the
expectation value of $\lambda_1$ and $\lambda_2$, as will be shown later.
\subsection{Expectation value of $\lambda_1$}
\vspace{0.5cm}
\subsubsection{Sub-critical regime, $\theta \ll 1$}
\begin{figure}
    \centering
    \begin{subfigure}{.5\textwidth}
        \centering
        \includegraphics[width=1.0\linewidth]{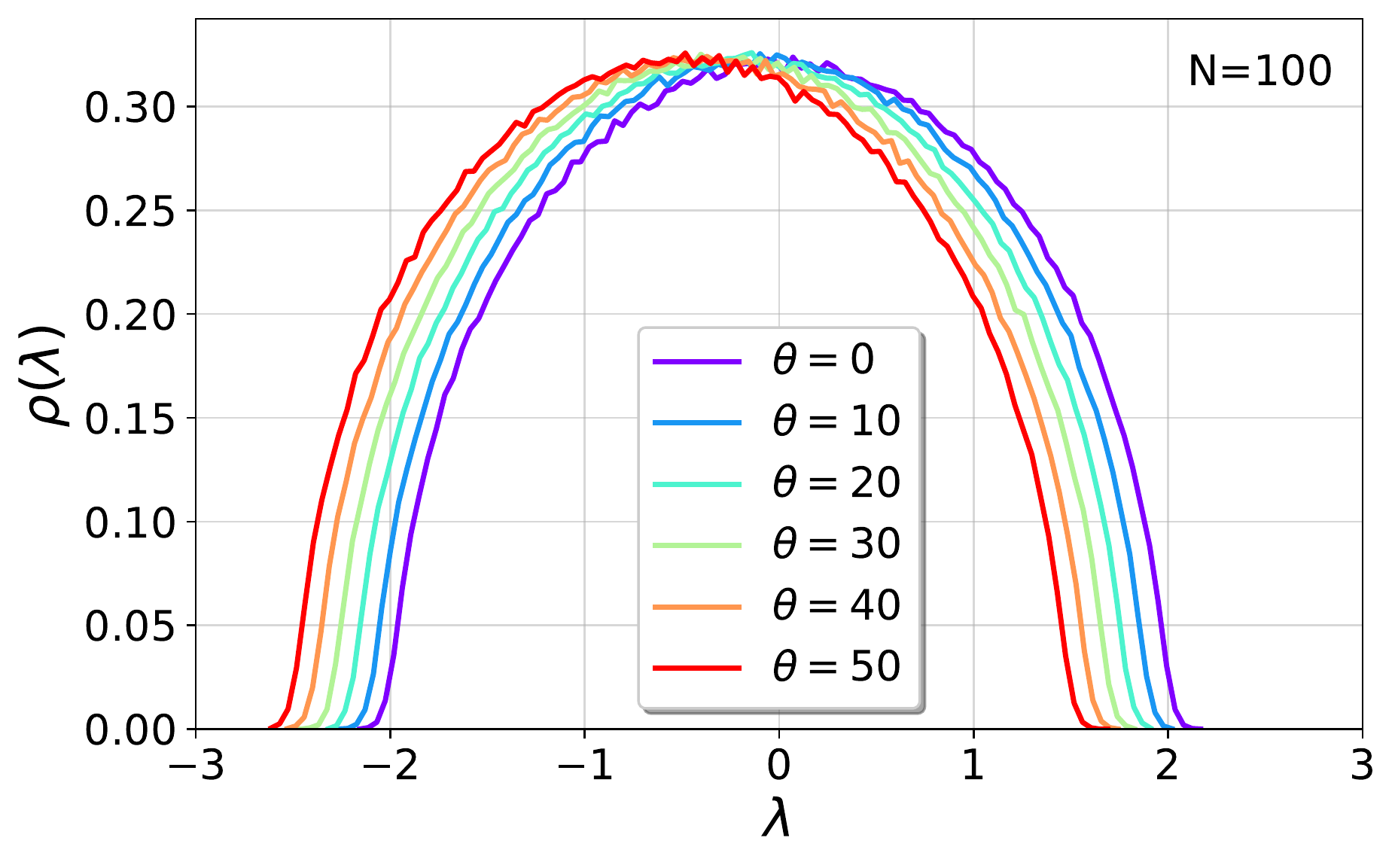}
        \caption{$N=100$}
        \label{fig:semicircle-shif-a}
    \end{subfigure}%
    \begin{subfigure}{.5\textwidth}
        \centering
        \includegraphics[width=1.0\linewidth]{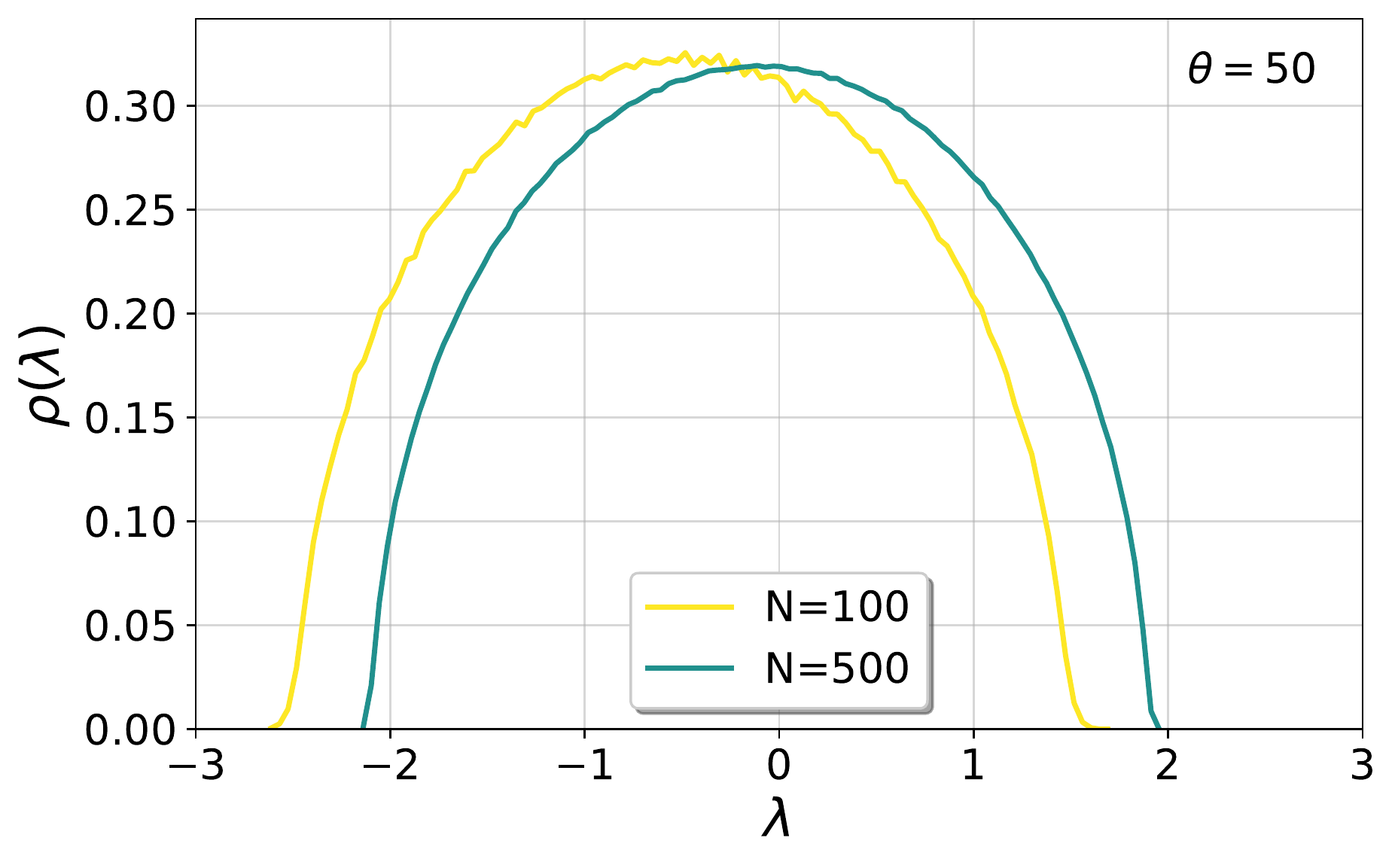}
        \caption{$\theta = 50$}
        \label{fig:semicircle-shif-b}
    \end{subfigure}
\caption{Shift of the semicircle as a consequence of the center of mass conservation described in the text.
(a) For fixed $N=100$, the semicircle moves to the left proportionally to the perturbation intensity.
(b) For fixed $\theta=50$, the shift depends on the size of the matrix: larger sizes $N$ suffer smaller shifts.}    
    \label{fig:semicircle-shift}
\end{figure}

\begin{figure}[h!]
\centering
\includegraphics[width=1\textwidth]{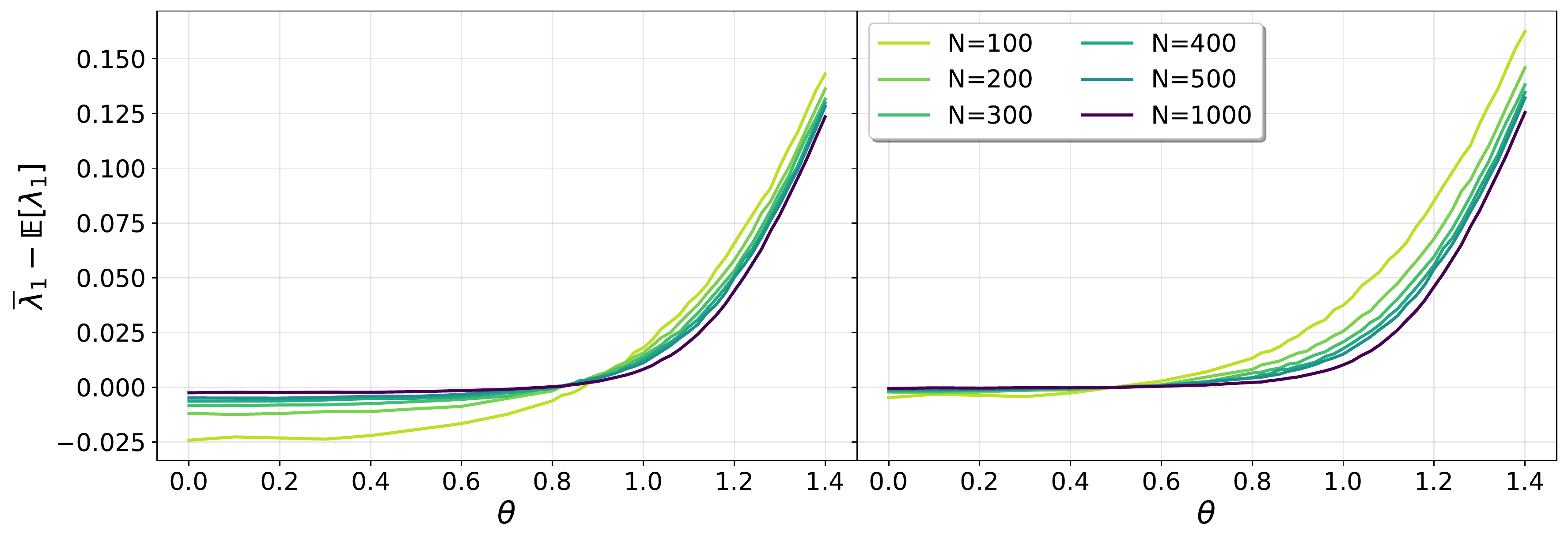}
\caption{Shift of the numerical average of $\lambda_1$ relative to the theoretical prediction
(\ref{eq:spike_l1_expval_pretrans}), for $\theta <1$. The left panel shows  results without considering the correction due
to the conservation of the center of mass. In the right panel, after inclusion of the correction, the data
shows a good collapse for growing sizes $N$ and sufficiently small values of $\theta$. }
\label{fig:spike_l1_av_pretrans}
\end{figure}

When $\theta < 1$ and for large but finite $N$, $\lambda_1$ is expected to behave as $\lambda_1 = 2 + \xi \,N^{-2/3}$,
where $\xi$ is a random variable described by the GOE TW distribution, with expected value $\mathbb{E}[\xi]=-1.21$
~\cite{TracyWidom1996,Majumdar2020}. Then, the expected value of $\lambda_1$ would behave as
$\mathbb{E}[\lambda_1]\approx2-1.21N^{-2/3}$.
Nevertheless, the previous result is valid when the diagonal elements of the random
matrix are non-null. In the present case the matrix $\boldsymbol{M}$ is traceless with all the diagonal elements equal
to zero. In Figure \ref{fig:semicircle-shif-a} we can see that the semicircle moves approximately linearly to the left
as the perturbation intensity increases, while in Figure \ref{fig:semicircle-shif-b} it is clear that bigger
matrix sizes $N$ suffer smaller shifts. 
This is a consequence of the traceless character of the matrix.
Upon changing the average value of its elements, $\theta/N$, the eigenvalues will have to rescale their expected
values in order to satisfy the condition that they must add up to zero.
This is analog to a center of mass conservation of the eigenvalue density.  
Then, because for finite $N$ the weight of each eigenvalue is $1/N$,
the expected value of $\lambda_1$ should approximately be given by:
\begin{equation}
\mathbb{E}[\lambda_1] \approx \left(2-1.21N^{-2/3}\right)\left(1-\frac{1}{N}\right),
\hspace{1cm} \theta \ll 1.
    \label{eq:spike_l1_expval_pretrans}
\end{equation}
Note that, because this correction acts equally on every eigenvalue, it will have no effect in the gap
$g=\lambda_1-\lambda_2$, which is the relevant quantity for the long time dynamics.
Figure \ref{fig:spike_l1_av_pretrans} shows the deviation of the numerical average of $\lambda_1$ from the
theoretical prediction (\ref{eq:spike_l1_expval_pretrans}). The improvement of the collapse after inclusion
of the center of mass conservation effect is evident in the right panel. It is also possible to note that the collapse
breaks down for $\theta >0.6$, for the sizes considered, when the system begins to cross over to the critical regime.
\vspace{0.5cm}
\subsubsection{Super-critical regime, $\theta \gg 1$}
\label{sec:l1-supercrit}
\begin{figure}[h!]
\centering
\includegraphics[width=1\textwidth]{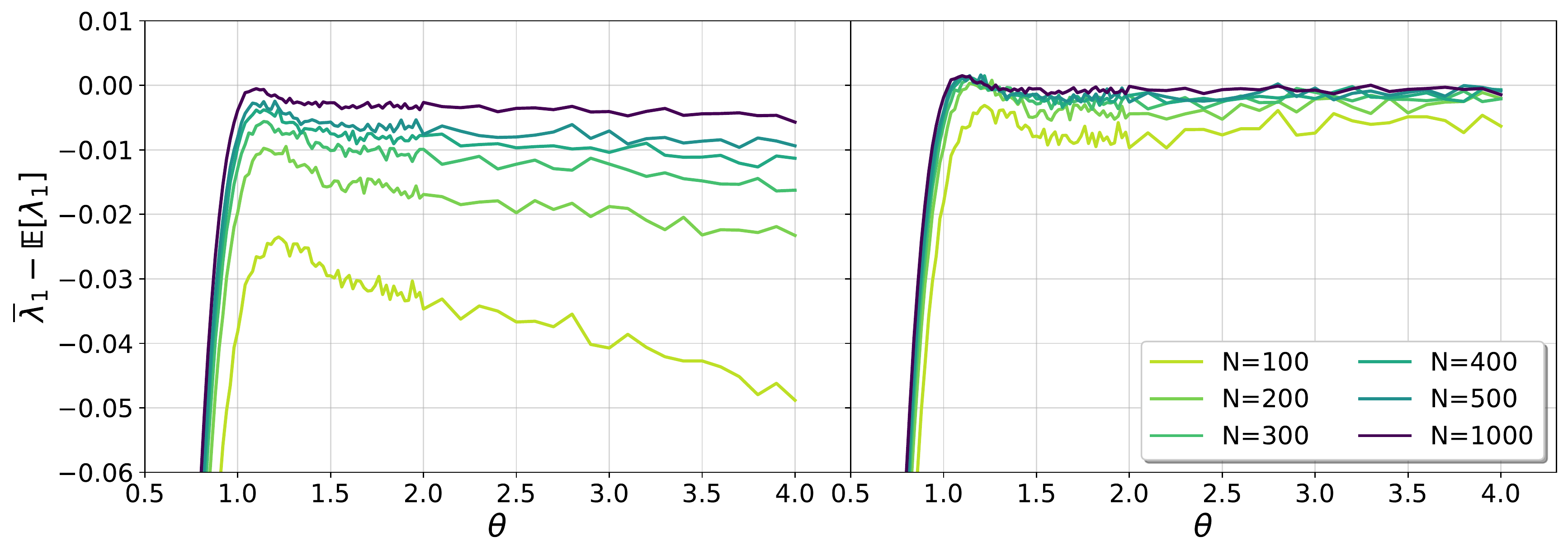}
\caption{Shift of the numerical average of $\lambda_1$ relative to the theoretical prediction
(\ref{eq:lambda1-average-theta-bigger1}), for $\theta > 1$. The left panel shows  results without
considering the correction due to the conservation of the center of mass. In the right panel,
after inclusion of the correction, the data shows a good collapse, improving as $N$ grows.}
\label{fig:spike_l1_av_postrans}
\end{figure}

This is the regime in which $\lambda_1$ becomes isolated from the bulk. In this case the finite $N$ fluctuations
are predicted to be Gaussian, given by equation (\ref{eq:spike_l1_gauss}). As in the $\theta <1$ case, the center of mass
conservation must be obeyed. We found that it amounts to a shift of the large $N$ result $(\theta+\theta^{-1})$
by the appropriate weigth factor $1/N$. Then, for $\theta \gg 1$, the expected value of $\lambda_1$ will be given by:
\begin{equation}
    \mathbb{E}[\lambda_1] \approx (\theta + \theta^{-1})
    \left(1-\frac{1}{N} \right), \hspace{2cm} \theta \gg 1.
    \label{eq:lambda1-average-theta-bigger1}
\end{equation}
In Figure \ref{fig:spike_l1_av_postrans} the effect of the center of mass correction for $\theta >1$ can be
appreciated. In this case, besides $N$, there is a dependence on $\theta$,
evident in the left panel of the figure. Upon considering the center of mass correction, the result
agrees well with equation (\ref{eq:lambda1-average-theta-bigger1}), when $\theta \gg 1$.
\vspace{0.5cm}
\subsubsection{Critical regime, $\theta \sim 1$}
\begin{figure}
    \centering
    \begin{subfigure}{.5\textwidth}
        \centering
        \includegraphics[width=1.0\linewidth]{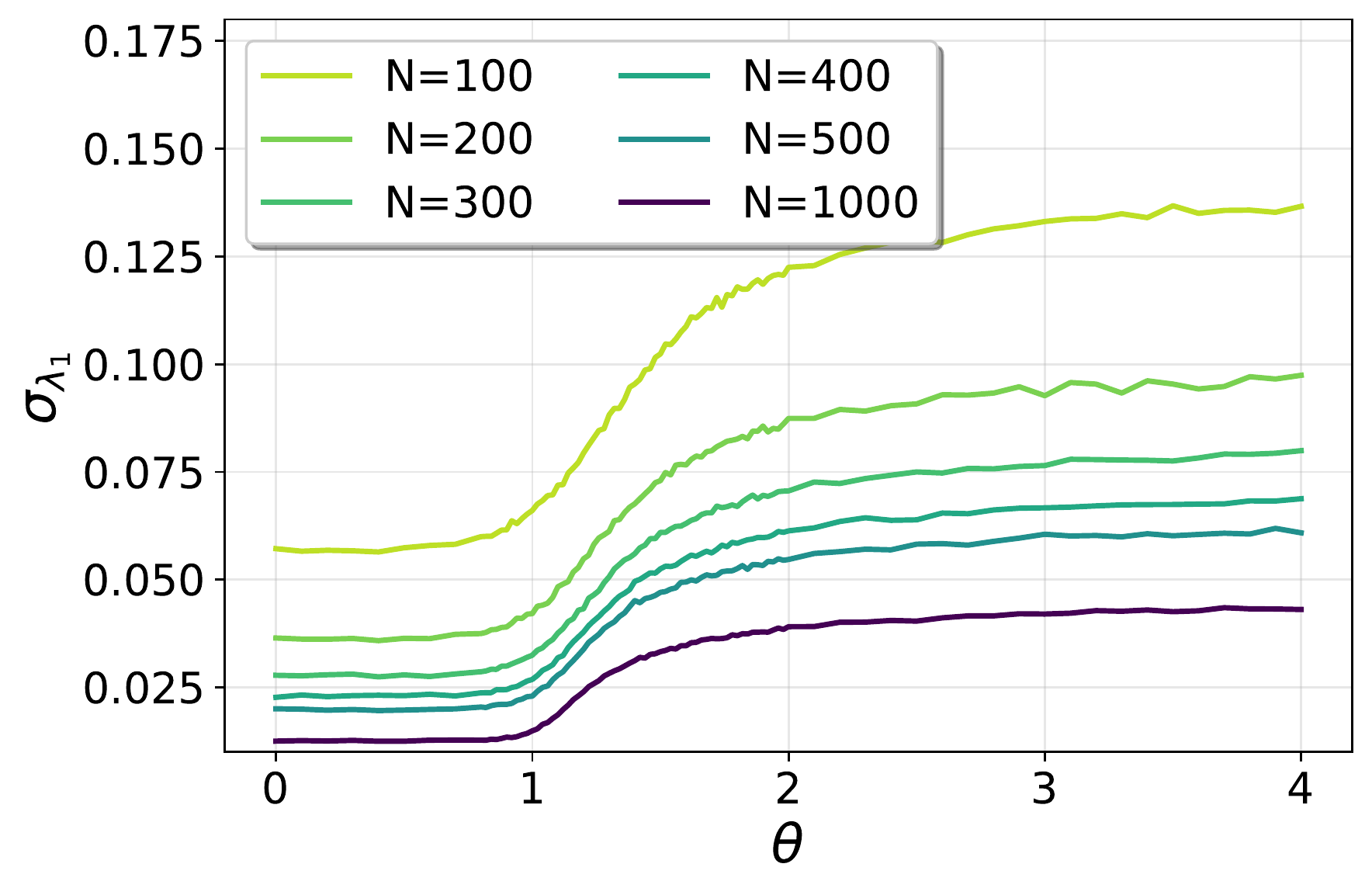}
        \label{fig:fluct_l1_crit-a}
    \end{subfigure}%
    \begin{subfigure}{.5\textwidth}
        \centering
        \includegraphics[width=1.0\linewidth]{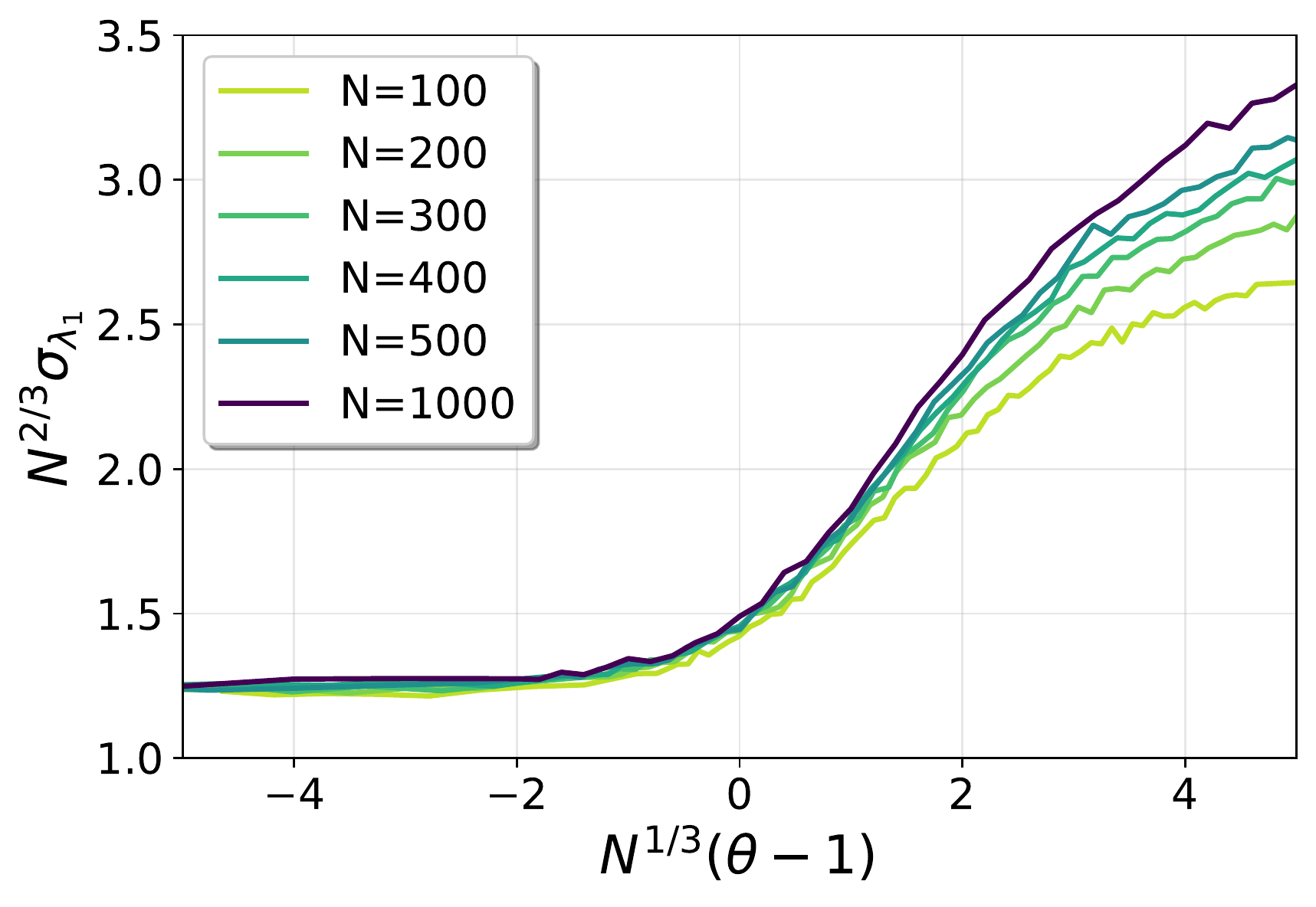}
        \label{fig:fluct_l1_crit-b}
    \end{subfigure}
\caption{Fluctuations of $\lambda_1$ in the critical regime.}
    \label{fig:fluct_l1_crit}
\end{figure}

In reference~\cite{Bloemental2013} the statistics of the largest eigenvalue of spike real Gaussian
random matrices in the critical regime is considered. The critical regime is defined for fixed values
of the parameter $\omega = N^{1/3}(\theta-1)\ \in (-\infty,\infty]$, and was described originally for
the spike complex Wishart ensemble in~\cite{Baik2005}, where the phenomenon of the phase transition
in the statistics of the largest eigenvalue was indentified. In Theorem 1.5 of~\cite{Bloemental2013} it is shown that,
in the critical regime, the eigenvalues of spike Gaussian random matrices are given in terms of the
eigenvalues of the {\em stochastic Airy operator} ${\cal H}_{\beta,\omega}$ with suitable boundary conditions.
For finite $\omega$, the statistics of the largest eigenvalue $\lambda_1$ is described by a
``one parameter family of deformations of the Tracy-Widom($\beta$)'' distributions, interpolating between the
usual values of $\beta=1,2,4$. As a consequence, in the critical regime, one expects the $\lambda_1$
fluctuations to be approximately described by the TW distribution, but not exactly, with a difference
that depends on the value of $\omega$. In Figure \ref{fig:fluct_l1_crit} we show the (numerical)
standard deviation of $\lambda_1$ for different system sizes. In the left panel the raw data is shown as
a function of $\theta$. In the right panel a data collapse is shown, with the scaling variable $\omega$
as defined above, assuming fluctuations to scale with $N^{2/3}$, as would be expected for a perfect Tracy-Widom
behavior. While the collapse is good for $\omega < 0$ and performs better in the whole interval as the
size $N$ grows, the quality of the collapse decays as $\omega$ grows. According to the results in
~\cite{Bloemental2013}, a continuous change in the exponent, away from $2/3$, should be expected.

\vspace{5cm}

\subsection{Expectation value of $\lambda_2$}
\begin{figure}[h!]
\centering
\includegraphics[width=1.0\linewidth,height=0.18\textheight]{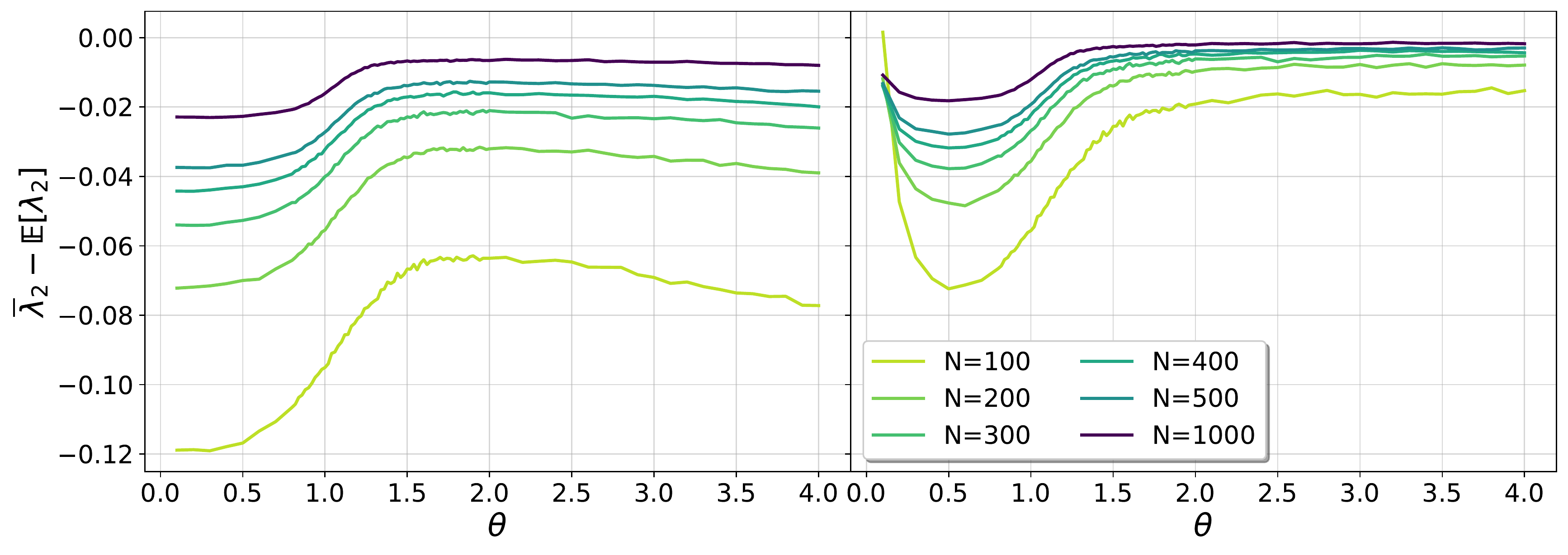}
\caption{Shift of the numerical average of $\lambda_2$ relative to the theoretical prediction
(\ref{eq:spike_l1_expval_pretrans}). The left panel shows  results without considering the correction due
to the conservation of the center of mass. In the right panel, after inclusion of the correction, the data
shows a good collapse for growing sizes $N$ and sufficiently large values of $\theta$.}
\label{fig:2ndEig-shift}
\end{figure}

\begin{figure}[h!]
\centering
\includegraphics[width=0.7\linewidth,height=0.25\textheight]{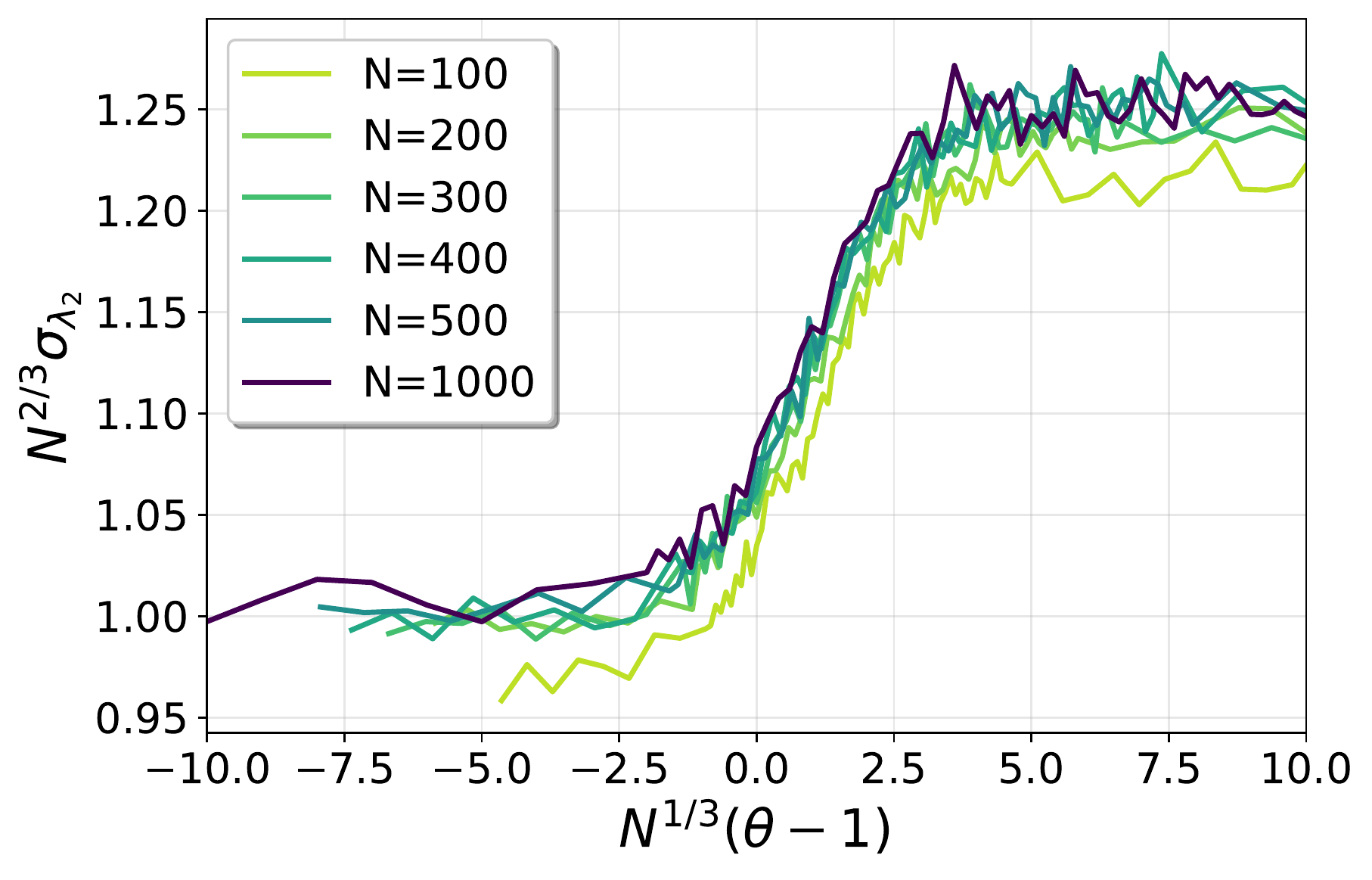}
\caption{Data collapse of the fluctuations of $\lambda_2$.}
\label{fig:fluc-lambda2-colapsed}
\end{figure}

As $\lambda_1$ jumps outside the semicircle of the Wigner law, it is expected that $\lambda_2$
will take its place at the soft edge of the eigenvalue density function. In particular, it is expected that $\lambda_2$
will show fluctuations given by the Tracy-Widom distribution. Then, the expectation value of $\lambda_2$
should behave as:
\begin{equation} \label{eq:meanlambda2}
    \mathbb{E}[\lambda_2]  \approx 
     (2 - 1.21N^{-2/3})\left(1-\frac{1}{N}\right) - \frac{(\theta+\theta^{-1})}{N}, \hspace{2cm} \theta \gg 1,
\end{equation}
where the first term corresponds to $\mathbb{E}[\lambda_1]$ in the sub-critical
regime, equation (\ref{eq:spike_l1_expval_pretrans}) and the second is the correction due to the center
of mass conservation when the largest eigenvalue has detached from the bulk.
The behavior of the shift of the numerical average $\overline{\lambda}_2$ from the expectation given by
eq. (\ref{eq:meanlambda2}) is shown in Figure \ref{fig:2ndEig-shift}. In the left panel
only the first term on the righthand side of (\ref{eq:meanlambda2}) is shown, while the right panel
shows the full expression, after taking into account the center of mass conservation term. A
progressive good collapse can be seen, in the $\theta \gg 1$ regime, as $N$ grows.
In Figure \ref{fig:fluc-lambda2-colapsed} we show a data collpase of the fluctuations of $\lambda_2$.
The collapse is good for the largest sizes, in agreement with theoretical expectations.

\subsection{Statistics of small gaps $g=\lambda_1-\lambda_2 \ll 1$}
\begin{figure}[h!]
\centering
\includegraphics[width=0.7\linewidth,height=0.25\textheight]{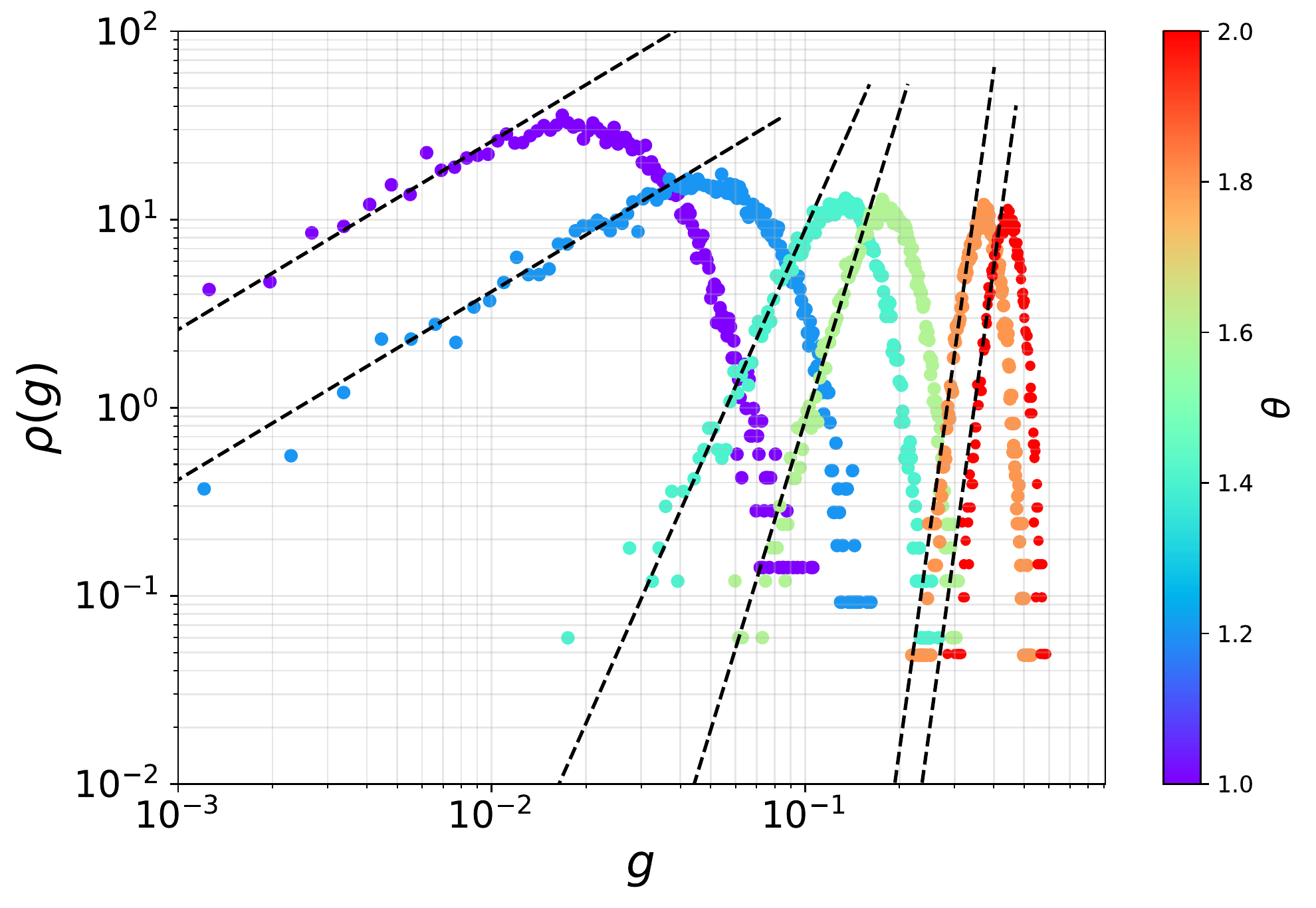}
\caption{Probability distribution functions of the gap between the two largest eigenvalues for an ensemble of
spike random matrices of size $N=1000$ and different values of the deterministic term intensity $\theta$. The
double log scale shows algebraic behavior at small $g$.}
    \label{fig:gapdist}
\end{figure}

\begin{figure}[h!]
\centering
\includegraphics[width=0.9\linewidth,height=0.2\textheight]{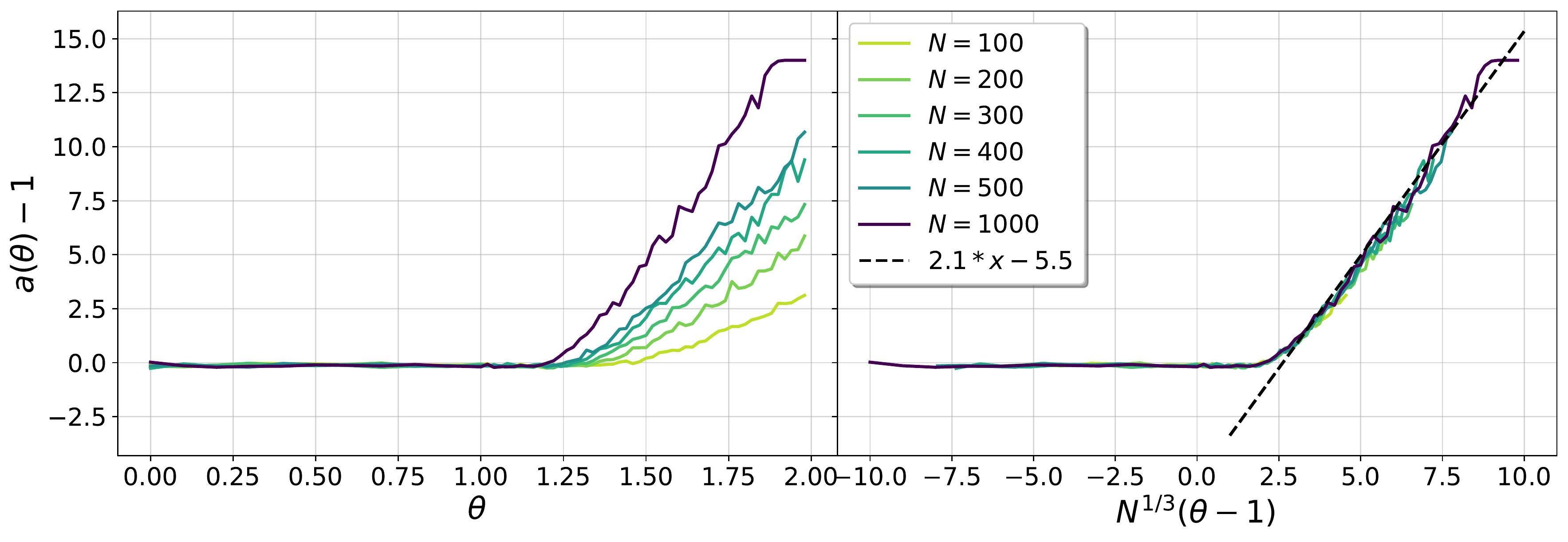}
\caption{The gap exponent $a(\theta,N)$.}
\label{fig:gap-param-A}
\end{figure}

The behavior of the gap between the two largest eigenvalues, $g=\lambda_1-\lambda_2$, will also depend
on the regime considered. In the sub-critical regime the effect of the deterministic
perturbation is negligible and one expects that the statistics of the gap will be governed by the
results of reference~\cite{PerretSchehr2015}. In turn, this will lead to power law time/size scalings, as
studied in~\cite{Fyodorov2015,Barbier2021}. In the super-critical regime, the two
largest eigenvalues become approximately independent random variables. From eq. (\ref{eq:spike_l1_gauss}),
$\lambda_1$ shows Gaussian flucutations which grow with $\theta$. Then, in this regime, the fluctuations of
the gap are expected to be Gaussian also, similar to what happens with $\lambda_1$. In turn, this will reflect in
exponential time relaxations of observables, like the energy gap, a typical behavior of
ferromagnetic phases. More interesting is the intermediate, critical regime. In this regime, in which 
$\lambda_1$ and $\lambda_2$ are strongly correlated, the statistical behavior of the gap is not known.
With the aim of describing the long time behavior of the energy gap, we have pursued a numerical characterization
of the statistics of the gap in the small gap regime, $g \ll 1$, relevant to the long time relaxation dynamics.
In Figure \ref{fig:gapdist} the distribution of the gap is shown in double logarithmic scale for an ensemble of
random spike matrices of size $N=1000$ and different values of $\theta \geq 1$. In all cases it can be seen that
the behavior is algebraic for small $g$.
Thus, for the small gaps regime, we expect that the pdf of the gap will approximately behave as:
\begin{equation}
    f(g) \sim b(\theta,N)\ g^{a(\theta,N)},
    \label{eq:gapsmallg}
\end{equation}
where $a(\theta,N)$ and $b(\theta,N)$ are parameters to be determined. We numerically adjusted those parameters
to fit the data in the interval $g < \mathbb{E}[g]-\sigma_g$. In Figure \ref{fig:gap-param-A} we show 
the behavior of the gap exponent with $\theta$, for different system sizes. The left panel shows that
$a(\theta,N)$ is a constant equal to one in the sub-critical regime, in agreement with the results for
the pure SSK model~\cite{PerretSchehr2015}.
For $\theta > 1$ the numerical analysis suggests a linear behavior, with a slope dependent with $N$.
A very good data collapse is obtained as a function of the critical scaling variable $N^{1/3}(\theta-1)$,
as can be seen in the right panel of Figure \ref{fig:gap-param-A}. Then, for $\theta > 1$,
the gap exponent behaves approximately as:
\begin{equation}
    a(\theta,N) \approx 1 + c_1\,N^{1/3}(\theta-1)-c_2,
    \label{eq:gap-param-a}
\end{equation}
where $c_1 \approx 2.1$ and $c_2 \approx 5.5$ are fit parameters.
With these results, now it is possible to compute the finite size behavior of the time dependent excess
energy, which is the subject of the next section.
\section{Long time decay of the excess energy} \label{sec:dynamics}
\subsection{$N \to \infty$ limit}
For random initial conditions given by (\ref{eq:rdminitcond}), the 
Lagrange multiplier is given by~\cite{Fyodorov2015,Barbier2021}:
\begin{equation}
z(t) = \frac{1}{2}\frac{d}{dt}\ln{\left[ \frac{1}{N}\sum_{\mu=1}^N e^{2\lambda_{\mu}t}\right]}
\xrightarrow{N\to\infty}
\frac{1}{2}\frac{d}{dt}\ln{\left[ \int_{-\infty}^{\infty}d\lambda \,\rho(\lambda)\,e^{2\lambda t}\right]},
\end{equation}
where the density of eigenvalues, $\rho(\lambda)$, is given by (\ref{eq:spike-spectrum}). Performing the
integrations, the exact solution is given by:
\begin{equation}
z(t) = \frac{I_1(4t)}{2t}+ e^{2(\theta+\theta^{-1})t}, \hspace{2cm} \theta > 1,
\end{equation}
where $I_1(x)$ is a modified Bessel function of the first kind. In the long time regime, the above
expression has the asymptotic behavior:
\begin{equation}
z(t) \to \left(\theta+\theta^{-1}\right) +
\left[2-\left(\theta+\theta^{-1}\right)\right]
\frac{e^{[4-2(\theta+\theta^{-1})]t}}{4\sqrt{2\pi}\,t^{3/2}}, \hspace{2cm} \theta > 1.
\end{equation}
Remembering that the Lagrange multiplier is proportional to the energy density, $z(t)=-2e(t)$ and that,
when $\theta >1$, the largest eigenvalue is given by $\lambda_1=\theta+\theta^{-1}$,
we obtain for the long time behavior of the average excess energy the result:
\begin{equation}
\lim_{N \to \infty}\mathbb{E}[\Delta e(t,\theta)] =
\begin{cases}
\varepsilon(t) & \text{for}\ \theta \leq 1\\
\frac{\left[\left(\theta+\theta^{-1}\right)-2\right]}{8\sqrt{2\pi}}
\frac{e^{[4-2(\theta+\theta^{-1})]t}}{t^{3/2}} & \text{for}\ \theta > 1,
\end{cases}
\label{eq:excess-spike-largeN}
\end{equation}
where $\varepsilon(t)=3/8t$ is the known result for the pure SSK model~\cite{Cugliandolo1995}. We note
that, as expected, in the $\theta >1$ ferromagnetic regime the asymptotic relaxation, when
$t \gg [4-2(\theta+\theta^{-1})]^{-1}$, is exponential,
faster than in the spin glass phase of the model.

\subsection{Finite system size}
With the results obtained for the pdf of the gap in the small gap regime for finite $N$,
we can compute the late time behavior of the average excess energy, as given by
eq. (\ref{eq:avexenergy}):
\begin{equation}
    \mathbb{E}[\Delta e(t,\theta,N)] = \int_0^{\infty} g\,e^{-2gt}\, f(g)\ dg,
\end{equation}
which can be decomposed in the form $\int_0^{\infty}=\int_0^r + \int_r^{\infty}$. In the long time
limit the dynamics will be dominated by the sector of small values of $g$. By a similar analysis to that
presented in equations (42)-(43) of ~\cite{Barbier2021}, it follows that the second integral will be negligible.
Then, using (\ref{eq:gapsmallg}):
\begin{equation}
    \mathbb{E}[\Delta e(t,\theta,N)] =  b(\theta,N) \int_0^{r} g\,e^{-2gt}\ g^{a(\theta,N)}\ dg,
\end{equation}
which has the exact solution:
\begin{equation}
    \mathbb{E}[\Delta e(t,\theta,N)] =  b(\theta,N)\  t^{-(2+a)}
    \left[ 2^{-(2+a)} \left(\Gamma(2+a)-\Gamma(2+a,2rt) \right)\right],
\end{equation}
with the limit $\lim_{rt \to \infty}\Gamma(2+a,2rt)=1$. From the analytical results and numerical analysis, 
we found that the behavior of the average excess energy, in the regime $\theta > 1$, can be described as follows:
\begin{equation}
    \mathbb{E}[\Delta e(t,\theta,N)]\sim
    \begin{cases}
        \upsilon(t)&\text{for}\;t < N^{\nicefrac{1}{3}} \\
        f_a(tN^{\nicefrac{-1}{3}})&\text{for} \;t > N^{\nicefrac{1}{3}}
    \end{cases}
\label{eq:excess-spike-fss}
\end{equation}
The function $\upsilon(t)$ represents the time relaxation of the excess energy in the thermodynamic
limit, given by (\ref{eq:excess-spike-largeN}) when $\theta >1$
and $f_a(x)=\nicefrac{c}{x^{2+a}}$ is a one parameter scaling function, with $a$ given by (\ref{eq:gap-param-a})
and $c$ a constant. It is to be noted the difference between the previous results and those for the
pure SSK model~\cite{Fyodorov2015,Barbier2021}. When $\theta >1$ and $t < N^{1/3}$ the relaxation is exponential,
given by $\upsilon(t)$, instead of the power law in the pure case. Also, the scaling exponent $1/3$ of the
algebraic regime differs from the $2/3$ of the pure SSK model.

\begin{figure}[ht!]
\centering
\includegraphics[width=1.0\linewidth]{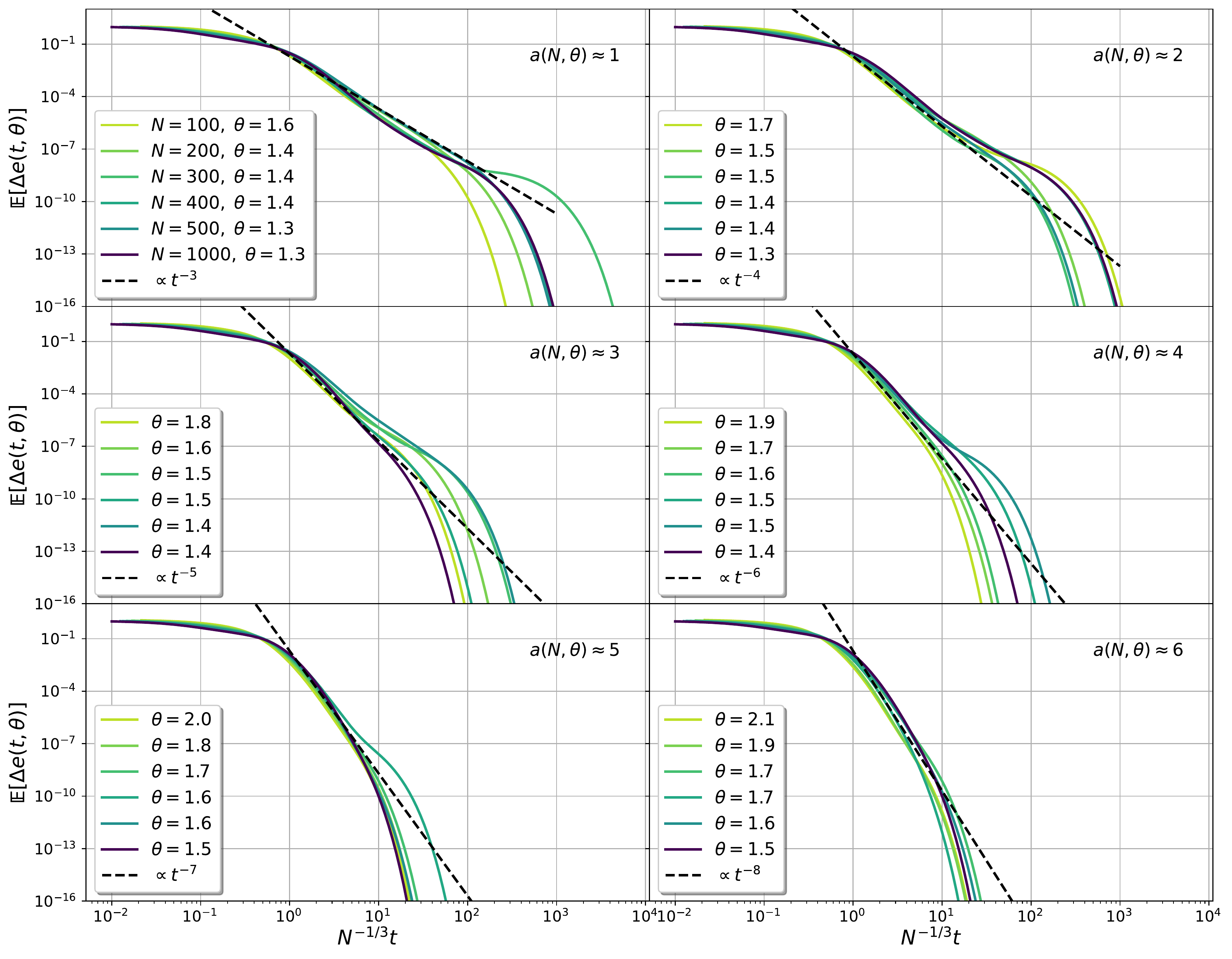}
\caption{Data collapse of the time decay of the average excess energy for the spike SSK model,
according to eq. (\ref{eq:excess-spike-fss}), for different values of the parameter $a(\theta,N)$.
The slope of the algebraic regime is governed by the one parameter scaling function
$f_a(x)$.}
\label{fig:energy-spike-collapse}
\end{figure}

In Fig. \ref{fig:energy-spike-collapse} we can see data collapses of the average excess energy of the spike
SSK model, for different fixed values of the parameter $a(\theta,N)$. They show good agreement with the previous
results. Similarly to the pure SSK model,
there is an algebraic scaling regime not present in the $N \to \infty$ limit. In this case, the exponent 
$2+a(\theta,N)$ depends on both the
spike intensity $\theta$ and the size of the system $N$, according to (\ref{eq:gap-param-a}).
In the figures, particular combinations of $\theta$ and $N$ were chosen in order to keep the
value of $a(\theta,N)$ approximately constant. As the exponent grows, so does the slope of the power law.
The relaxation becomes faster, but a power law regime can be identified by values of the parameter $a$ as
large as $a=7$, as shown in the figure.

\vspace{3cm}

\section{Discussion and conclusions}
\label{sec:conclusions}
In the first part of this work, we have presented a numerical study of the statistics of the two largest
eigenvalues and the gap, for random
matrices from the Gaussian Orthogonal Ensemble perturbed by a deterministic rank one matrix. 
The largest eigenvalue of such spike random matrices is known to go through a phase transition as the
intensity of the deterministic perturbation attains a critical value~\cite{Baik2005,Bloemental2013}.
The statistics in the sub-critical and super-critical
regimes are well described in the literature while results for the critical regime are scarce
~\cite{Mo2012,Bloemental2013}.
Our numerical analysis on the average values of the two largest eigenvalues confirmed analytical results from the
literature, after inclusion of additional size effects due to the traceless character of the matrices considered in this
work, which are of interest in physics models. In the critical regime, we showed results on the fluctuations of
$\lambda_1$ and $\lambda_2$ in good agreement with available theoretical predictions on
the existence of a critical scaling regime where the fluctuations are described by a one parameter family of scaling
functions, which can be seen as continuous deformations of the Tracy-Widom distribution~\cite{Mo2012,Bloemental2013}.
While our results are
compatible with that conclusions, more work is needed to extend and clarify the interpretation of the results
from the mathematical literature in the physical models context.

For the statistics of the gap, at present there are no known analytical results.
Then, we pursued a numerical characterization of the small gap regime of the probability density function,
which is the relevant regime for the long time behavior of physical observables. We show evidence that the pdf
of the gap has a power law behavior for small gaps. The exponent of the power law depends on both
the intensity of the deterministic perturbation $\theta$ and the system size $N$, in the form which defines the
critical sector of the model, $a(\theta,N) \sim N^{1/3}(\theta-1)$, when $\theta >1$. 
After the
characterization of the pdf of the gap, we described the long time decay of the average excess energy
of the Spherical Sherrington-Kirkpatrick model with a Curie-Weiss perturbation term.
We first obtained the analytical result in the large $N$ limit, showing that, as expected, the relaxation is
exponential for $\theta >1$. We then considered the large but finite $N$ behavior.
The most interesting and new regime
to describe is near the phase
transtion between the spin glass and ferromagnetic phases. In this critical regime, using the results obtained for
the gap pdf, we showed the existence of a
sector with power law relaxation as a function of the scaling variable $tN^{-1/3}$.

The results for the gap pdf and the excess energy relaxation are our main new results, not previously reported in the
literature. Being mainly of a numerical character, we expect that they will motivate to pursue analytical approaches to
the computation of the gap probability distribution function, which has been shown to be a relevant random variable
to describe the late time dynamics of spherical models with pairwise interactions.

\acknowledgments
The research of P.H.F.P. was funded in part by Fundação de Amparo à Pesquisa do Estado do Rio de Janeiro (FAPERJ).
D.A.S. was funded in part by CNPq through a research fellowship.


%

\end{document}